\documentclass[reprint,amsmath,amssymb,aps,prd,longbibliography,preprintnumbers,showpacs,floatfix,
nofootinbib]{revtex4-1}

\usepackage{color,graphicx} 
\usepackage{dsfont}

\begin{document}

\preprint{DESY 11-164}
\title{Tetraquark Interpretation of the Charged Bottomonium-like states $Z_b^\pm(10610)$ and
 $Z_b^\pm(10650)$ and Implications}

\author{Ahmed~Ali}
\email{ahmed.ali@desy.de}

\author{Christian Hambrock}
\email{christian.hambrock@desy.de}

\author{Wei~Wang}
\email{wei.wang@desy.de}
\affiliation{Deutsches Elektronen-Synchrotron DESY, D-22607 Hamburg, Germany}

\begin{abstract}
We present a tetraquark interpretation of the charged bottomonium-like states $Z^\pm_b(10610)$ and
$Z^\pm_b(10650)$,
observed by the Belle collaboration in the $\pi^\pm \Upsilon(nS)$ $(n=1,2,3)$ and
$\pi^\pm h_b(mP)$ $(m=1,2)$ invariant mass spectra from
the data taken 
near the peak of the $\Upsilon(5S)$.
In this framework, the underlying processes involve the
production and decays of a vector tetraquark  $Y_b(10890)$,
 $e^+e^- \to Y_b(10890) \to [Z_b^\pm (10610)\pi^\mp, Z_b^\pm (10650)\pi^\mp]$ followed by the
decays $[Z_b^\pm (10610),  Z_b^\pm (10650)] \to \pi^\pm \Upsilon(nS), \pi^\pm h_b(mP)$.
Combining the contributions from the meson loops and  an effective Hamiltonian,
 we are able to reproduce the observed
 masses of the $Z_b^\pm (10610)$ and $Z_b^\pm (10650)$. 
The analysis presented here is in agreement with the Belle data and  provides
crucial tests of the tetraquark hypothesis. We also calculate the corresponding meson loop effects 
in the charm sector and find them dynamically suppressed. The charged charmonium-like states
$Z_c^\pm(3752)$ and $Z_c^\pm(3882)$ can be searched for in the decays of the $J^{PC}=1^{--}$
tetraquark state $Y(4260)$ via $Y(4260) \to Z_c^\pm(3752)\pi^\mp$ and 
$Y(4260) \to Z_c^\pm(3882)\pi^\mp$,
with the subsequent decays $(Z_c^\pm(3752),Z_c^\pm(3882))  \to (J/\psi, h_c)\pi^\pm$. 

\end{abstract}

\maketitle

Recently Belle~\cite{Collaboration:2011pd} (updating a previous publication~\cite{Collaboration:2011gj}) reported the measurement of the
 $\pi^\pm \Upsilon(nS)(n=1,2,3)$ and $\pi^\pm h_b(mP) (m=1,2)$ invariant mass spectra
from the data taken  near the peak
 of  the $\Upsilon (5S)$ resonance in the processes $e^+e^-\to \Upsilon(nS) \pi^+\pi^-$
 and $ e^+e^-\to h_b(mP)\pi^+\pi^-$, in which two charged bottomonium-like states
 $Z^\pm_b(10610)$ and $Z^\pm_b(10650)$ are discovered. Hereafter, these states will be abbreviated
 to $Z_b$ and $Z_b^\prime$, respectively.
The masses and decay widths averaged over the five different final states
 are in MeV~\cite{Collaboration:2011pd}:
 \begin{equation}
\begin{array}{cccccc}
m_{Z^\pm_b}&=&10607.2\pm2.0\, 
,&
m_{Z'^\pm_b}&=&10652.2\pm1.5\, 
,\nonumber\\
\Gamma_{Z^\pm_b}&=&18.4\pm2.4\, 
,&
\Gamma_{Z'^\pm_b}&=&11.5\pm2.2\, 
.\label{eq:belledata}
\end{array}
 \end{equation}
The angular distribution analysis indicates that the
quantum numbers of both $Z^\pm_b$ and $Z'^\pm_b$ are
$I^{G}(J^P)=1^+(1^+)$.  
These states defy a standard bottomonium assignment,
as in the valence approximation they consist of four  quarks $bu\bar{b} \bar{d}$
 (and charge conjugates). 
 
Due to the proximity of the  $Z_b$ and $Z_b^\prime$ masses with the $B\bar{B}^*$ and $B^*\bar{B}^*$ thresholds~\cite{Nakamura:2010zzi}, it has been proposed that the former  
could be realized as $S$-wave $B\bar{B}^*$ and
$B^*\bar{B}^*$  molecular states, respectively~\cite{Bondar:2011ev,Voloshin:2011qa,Zhang:2011jja,Yang:2011rp,Sun:2011uh,Cleven:2011gp,Mehen:2011yh}. In this scenario, the heavy quark spin
structure of the $Z_b$ and $Z_b^\prime$ is expected to mimic that of the corresponding
meson pairs 
\begin{eqnarray}
| Z_b^\prime\rangle
&=& (0^-_{b\bar b}\otimes 1^-_{q\bar q} -1^-_{b\bar b}\otimes 0^-_{q\bar q})/\sqrt 2,\nonumber\\
 | Z_b\rangle
&=&  (0^-_{b\bar b}\otimes 1^-_{q\bar q} +1^-_{b\bar b}\otimes 0^-_{q\bar q})/\sqrt 2,
 \label{eq:Fierz-Bondar}
\end{eqnarray} 
where $0^-$ and $1^-$ denotes the para and ortho- states with negative parity, respectively.
One anticipates 
the mass splitting to follow $\Delta m_{Z_b} \equiv  m_{Z_b^\prime}-m_{Z_b}= m_{B^*}-m_{B} \simeq 46$ MeV,
in neat agreement with the observed value $\Delta m_{Z_b} = (45\pm 2.5)$ MeV~\cite{Collaboration:2011pd}. Moreover,
the structure in Eq.~(\ref{eq:Fierz-Bondar})  predicts that  $Z_b$ and $Z_b^\prime$
 should have 
the same decay width, which is  approximately in agreement with the data.

Despite these striking patterns, the fact that both $Z_b$
and $Z_b^\prime$ lie nominally above their respective thresholds by about 2 MeV
 reveals a tension with the molecular interpretation. If consolidated by more precise
experiments, this feature may become a
serious problem in this approach, as a one-pion exchange potential, which would
 produce such a bound state, does not support an $S$-wave
$B \bar{B}^*$ resonance above threshold in an effective field theory~\cite{Nieves:2011vw}.
Also, the measured total decay widths appear much too large compared to the 
naively expected ones for loosely bound states, and this suggests that both $Z_b$ and $Z_b^\prime$
are compact hadrons. In addition, 
the measured cross sections in question are too big
to be interpreted  in terms of the decays $\Upsilon(5S) \to (\Upsilon(nS), h_b(mP)) \pi^+\pi^-$. 

In this paper, we pursue a different ansatz in which the observed processes arise
from the production and decays of  a vector tetraquark 
$Y_b(10890)$~\cite{Ali:2009pi,Ali:2009es,Ali:2010pq},
 having a (Breit-Wigner) resonant mass of
$[10888.4 ^{+2.7}_{-2.6} ({\rm stat}) \pm 1.2 ({\rm syst})]$ MeV and a width of
$[30.7 ^{+8.3}_{-7.0}({\rm stat}) \pm 3.1 ({\rm syst})]$ MeV
~\cite{Abe:2007tk,Chen:2008xia}. The mass and, in particular, the decay width of
 $Y_b(10890)$ differ from the
 Particle Data Group entries assigned to the $\Upsilon(5S)$~\cite{Nakamura:2010zzi}. 
We propose that the states $Z_b$ and $Z_b^\prime$ seen in the decays of 
$Y_b(10890)$ are themselves charged tetraquark candidates having the
flavor configuration $[bu][\bar{b}\bar{d}]$ (and charge conjugates)
(see Refs.~\cite{Guo:2011gu,Cui:2011fj} for earlier suggestions along these lines).
  Their neutral isospin counterparts with $I_3=0$ have 
$J^{PC}=1^{+-}$ and their masses were calculated in the effective Hamiltonian
approach in~\cite{Ali:2009pi}. Ignoring the small isospin-breaking
effects~\cite{Ali:2009pi,Maiani:2004vq},  $Z_b$ and $Z_b^\prime$ have  the same
masses as those of their neutral counterparts. As shown below, these estimates
yield a too large value for $\Delta m_{Z_b}$ compared to the Belle measurement.

However, threshold effects and common decay channels may play an important role
beyond what can be described by the constituent quark model and its transcribed
form adopted in \cite{Maiani:2004vq} to work out the spectroscopic aspects for the
tetraquark states. In particular, two hadronic states having the same quantum 
numbers may mix due to dynamical effects, leading to differences in their masses
and decay widths. Typically, the resulting mass shift is dominated by decays to
the common states and reflects the partial widths to these states.  
 A case in point here is the mass difference between
the $D^0$ and $\bar{D}^0$, which is dominated by such common decay channels,
and is usually calculated by the meson loops, as dictated by
the optical theorem~\cite{Donoghue:1985hh}.
 Following essentially the same
line of argument,  we quantify this effect  for the two charged-bottomonium-like
states $Z^\pm_b$ and $Z'^\pm_b$. We recalculate the masses of  $Z_b$ and
 $Z_b^\prime$ states by taking into account the  meson loop contributions
 involving the Zweig-allowed two-body 
intermediate states $B \bar{B}^*$, $B^* \bar{B}^*$, $h_b(mP)\pi$, $\Upsilon(nS)\pi$
and $\eta_b\rho$. 
Theoretical estimates presented here account for the observed masses; in particular,
the precisely measured mass difference $\Delta m_{Z_b}$ is reproduced in terms of the 
partial decay widths of   $Z_b$ and $Z_b^\prime$. This can be tested in future  when the partial decay widths are measured precisely.  
In our approach, the mass eigenstates
$Z_b$ and $Z_b^\prime$ are rotated with respect to the tetraquark spin states
 $\tilde Z_b$ and $\tilde Z_b^\prime$, and we determine this mixing angle.

 We also work out the corresponding meson loop effects 
for the charged charmonium-like states $Z^\pm_c$ and $Z'^\pm_c$, with each one of them
belonging to an isotriplet. The masses of the electrically neutral states have been
calculated and are predicted to have typical values $m(Z_c)=3752$ MeV and
$m(Z^{\prime}_c)=3882$ MeV~\cite{Maiani:2004vq}. Ignoring the small isospin-breaking
 effects, these estimates apply for the  
charged counterparts   $Z^\pm_c$ and $Z'^\pm_c$ as well. We find that the meson-loop
effects are in this case dynamically suppressed, as detailed below. However,
 we do expect that the
production and decays of the $Z^\pm_c$ and $Z'^\pm_c$ will essentially mimic the
patterns seen for their bottomonium counterparts $Z^\pm_b$ and $Z'^\pm_b$. In
 particular, $Z^\pm_c$ and $Z'^\pm_c$, which are not measured so far, can be
 searched for in the decays of the
neutral $J^{PC}=1^{--}$ state $Y(4260)$, via the processes  
$Y(4260) \to Z_c^\pm(3752)\pi^\mp$ and 
$Y(4260) \to Z_c^\pm(3882)\pi^\mp$,
with the subsequent decays $(Z_c^\pm(3752),Z_c^\pm(3882))  \to (J/\psi, h_c)\pi^\pm$.

We  start with the classification of the $\tilde Z_b$ and $\tilde Z_b^\prime$ 
tetraquark states  in terms of the  spin and orbital angular momentum of the constituent diquark
and antidiquark. The concept of diquark was introduced by Gell-mann in his
epochal paper on quarks~\cite{GellMann:1964nj} and since then has been widely
discussed in the literature (for reviews on diquarks, see Refs.~\cite{Anselmino:1992vg,Jaffe:2004ph}). A diquark has positive parity and may be a scalar (spin-0, or ``good'' diquark) or an
 axial-vector (spin-1, or ``bad'' diquark)~\cite{Jaffe:1976ih,Jaffe:1978bu,Jaffe:2003sg} and is assumed to be a color
 antitriplet $\overline{3}_c$. 
The states $\tilde Z_b$ and $\tilde Z_b^\prime$ arise from the production and
decays of $Y_b(10890)$, identified with a linear combination of the two tetraquark states 
$Y_{[bq]}=[bq][\bar{b}\bar{q}]$ ($q=u,d$) 
having the spin and orbital momentum quantum numbers:
$S_{[bq]}=0$, $S_{[\bar{b}\bar{q}]}=0$,
$S_{[bq][\bar{b}\bar{q}]}=0$,  $L_{[bq][\bar{b}\bar{q}]}=1$, 
with the total spin
 $J_{[bq][\bar{b}\bar{q}]}=1$~\cite{Ali:2009pi}.
 We shall be using a non-relativistic notation to characterize the tetraquark states
 $|S_{[bq]}, S_{[\bar{b}\bar{q}]}; J_{{[bq]}[\bar{b}\bar{q}]}\rangle$,
 in which a matrix representation of
 the interpolating operators is used in terms of the $2\times 2$ Pauli matrices
 $\sigma_i$ $(i=1,2,3)$:
 $0_{[QQ]}\equiv Q^T\sigma_{2} Q/\sqrt{2}$, $1_{[QQ]}\equiv Q^T\sigma_{2}\sigma^{i} Q/\sqrt{2}$ and 
$0_{Q\bar Q}\equiv \bar Q  Q/\sqrt{2}$, $1_{Q\bar Q}\equiv \bar Q\sigma ^{i}  Q/\sqrt{2}$, $Q$ being any quark. The two tetraquark spin states $\tilde Z_b$ and $\tilde Z_b^\prime$  are represented as
\begin{eqnarray}
|\tilde Z_b\rangle&=& \big(0_{[bq]}\otimes 1_{[\bar{b}\bar{q}]} -1_{[bq]}\otimes 0_{[\bar{b}\bar{q}]}\big)/\sqrt 2,\nonumber\\ 
|\tilde Z_b^\prime\rangle&=& 1_{[bq]}\otimes 1_{ [\bar b\bar q]}.
\end{eqnarray}
Performing a Fierz transformation, the flavor and spin content in the 
 $b\bar b\otimes q\bar q$ and $b\bar q\otimes q\bar b$ product space can be made explicit:
\begin{eqnarray}
|\tilde Z_b\rangle
&=&   (- 1^-_{b\bar b}\otimes  0^-_{q\bar q} +0^-_{b\bar b}\otimes 1^-_{q\bar q} )/\sqrt 2 = 1^-_{b\bar q}\otimes 1^-_{q\bar b},\nonumber\\
|\tilde Z_b^\prime\rangle
&=& ( 1^-_{b\bar b}\otimes  0^-_{q\bar q} +0^-_{b\bar b}\otimes 1^-_{q\bar q})/\sqrt 2\nonumber\\
&=& ( 1^-_{b\bar q}\otimes 0^-_{q\bar b}+0^-_{b\bar q}\otimes 1^-_{q\bar b})/\sqrt 2.
\label{eq:Fierz}
\end{eqnarray} 
Eq.~(\ref{eq:Fierz}) shows that the $\tilde Z_b$ and $\tilde Z_b'$  have
similar coupling strengths with different final states.
The labels $0_{b\bar q}$ and $1_{b\bar q}$ 
in  Eq.~(\ref{eq:Fierz}) can be viewed as $\bar B$ and $\bar B^*$,
respectively. It follows that
$\tilde Z_b$  couples to $B^*\bar B^*$
state  while $\tilde Z_b^\prime$ couples to $B\bar B^*$.
We stress that this identification is in contrast with the molecular interpretation, in which
the transition  $\tilde Z_b^\prime\to B\bar B^*+h.c.$ is forbidden by the spin symmetry since
$\tilde Z_b^\prime$ is assumed to be essentially 
a  $B^*\bar B^*$ molecule~\cite{Bondar:2011ev}. 
 This difference can be tested in the future 
and  is of great importance in order to distinguish 
between the tetraquark and the hadronic molecule interpretations.

In the effective Hamiltonian approach, 
the $2\times 2$ mass matrix for the $S$-wave $1^{+}$ tetraquarks
$\hat M$ is given by~\cite{Maiani:2004vq}
\begin{eqnarray}
\hat  M
&=& \left(
2m_{[bq]} +\frac{3}{2} \Delta - \frac{\kappa_{q\bar q}+ \kappa_{b\bar b} }{2}
\right)\mathbb{I}+\left(\begin{array}{cc}
                                           -a & b \\
                                           b & a
                                          \end{array}\right)~,
\label{eq:massmatrix}
\end{eqnarray}
where $\mathbb{I}$ is a $2\times 2$ unit matrix, $a= \Delta /2 +(\kappa_{bq})_{\bar 3}- \kappa_{b\bar q}  $ and $b= \kappa_{q\bar q}-\kappa_{b\bar b}$.
In the above $(\kappa_{bq})_{\bar 3}$ accounts for the spin-spin interaction between
the quarks inside the diquark and antidiquark, $\kappa_{q\bar q}$ and $\kappa_{b\bar b}$ are
the couplings accounting for the interaction between the the quarks in the diquark to the antiquarks in the antidiquark,
and  $\Delta$ is the mass difference between the spin-1 and spin-0  diquarks.
Using the default values of the parameters~\cite{Ali:2009pi,Ali:2009es} (in units of MeV) 
\begin{eqnarray}
 \kappa_{q\bar q}= 79.5,\;\;
\kappa_{b\bar b}= 9,\;\;\; \kappa_{b\bar q}= 5.75, \;\; (\kappa_{bq})_3= 6,
\end{eqnarray}
yields the diquark mass $m_{[bq]}\simeq 5200$ MeV (from the $Y_b(10890)$ mass). The value of 
 $\Delta$ is uncertain,
with $\Delta \simeq 200$ MeV for the light quarks~\cite {Jaffe:2003sg}.
Reducing its value drastically  for the $c$ and $b$ quarks will reduce the level spacing of the
corresponding tetraquark states for which the experimental evidence is
rather sparse. Due to the lack of data, we adopt an admittedly somewhat arbitrary
 range  
$\Delta= (120\pm 30) $ MeV for our numerical calculations.  These parameters 
yield  the following values for the two charged tetraquark masses and the mass difference 
\begin{eqnarray}
 m_{Z_b}=(10443^{+35}_{-36}) {\rm MeV}, \; m_{Z_b^\prime}= (10628^{+53}_{-54}) {\rm MeV},
 \nonumber\\
\Delta m_{Z_b}=2\sqrt {a^2+b^2}= (185^{+21}_{-18}) {\rm MeV}.
\end{eqnarray}

We note that the prediction for $\Delta m_{Z_b}$ given above is much
larger than the experimental data, and there is no easy-fix
for this mismatch at hand in terms of the parameters in the effective Hamiltonian.
Since this Hamiltonian~\cite{Maiani:2004vq} adequately describes the mass spectrum 
of the exotic states discovered in the charm sector, we continue to use this as
our starting point and  argue that additional dynamical contributions to the
mass matrix arise from  the meson loops.

\begin{figure}
\begin{center}
\includegraphics[scale=0.4]{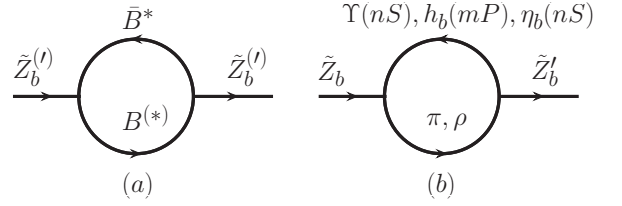}
\caption{Two-body meson-loop corrections to the function $\Sigma(s)$ defined in
Eq.~(\ref{eq:sigmas}).
 The intermediate states
  $B^{(*)}\bar B^*$ contribute only to the diagonal terms in the mass matrix while 
 $\Upsilon(nS)\pi$, $h_b(mP)\pi$ and
$\eta_b(nS)\rho$ contribute to both the diagonal and non-diagonal terms.
} \label{fig:mesonloop}
\end{center}
\end{figure}

With this premise, the renormalized masses can be obtained by computing
the two-point functions. 
At the one-loop level, the self-energy corrections
to the unperturbed propagator $\Sigma(p^2)g_{\mu\nu}$, depicted in Fig.~\ref{fig:mesonloop},
are written as 
\begin{eqnarray}
 \frac{-i(g^{\mu\alpha}- {p^\mu p^\alpha}/{p^2})}{p^2-\hat M^2}  i \Sigma(p^2) g_{\alpha\beta} 
 \frac{-i(g^{\beta\nu}- {p^\beta p^\nu}/{p^2})}{p^2-\hat M^2}.
\end{eqnarray}
Taking the $h_b\pi$ state as an example, we find 
\begin{eqnarray}
 \Sigma(s)  &=&    
\frac{g_{\tilde Z_b^{(\prime)} h_b\pi}g_{\tilde Z_b^{(\prime)} h_b\pi}^*}{(4\pi)^2}
\int_0^1 dx s \Lambda \Big[1-  \log\big(\frac{\Lambda}{\mu^2}\big)\Big],
\label{eq:sigmas}
\end{eqnarray}
where $\Lambda= x^2 s-xs +xm_\pi^2 +(1-x) m_{h_b}^2-i\epsilon$, and  the
coupling constants appearing above are defined through the hadronic
 interaction
\begin{eqnarray}
 {\cal L}= \epsilon_{\mu\nu\alpha\beta} g_{\tilde Z_b^{(\prime)}h_b\pi} \partial^\mu \tilde Z_b^{(\prime)\nu} \partial^\alpha h_b^\beta \pi+ h.c..\label{eq:hbLagrangian}
\end{eqnarray}
In deriving $\Sigma(s)$, the $\overline {\rm MS}$ scheme 
in the unitarity gauge 
has been used to remove the UV divergence with the scale $\mu\sim m_{Z_b^{(\prime)}}$.
We recall that the real parts of $\Sigma(s)$  contribute to the mass matrix,
 while the imaginary parts of $\Sigma(s)$ are related to the decay widths of $\tilde Z_b$ and $\tilde Z_b^\prime$.
In particular, the transitions $\tilde Z_b\to (\Upsilon(nS)\pi, h_{b}(mP)\pi,\eta_b(nS)\rho) \to \tilde Z_b^\prime$
contribute to the off-diagonal terms in the $2\times 2$ mass matrix and provide significant effects
on the mixing of the two tetraquark-spin eigenstates.

The meson loop corrections due to the different hadronic channels  can be viewed  as
 renormalizing the ``bare'' mass for the states predicted in the constituent
 tetraquark model. We are interested in the relative mass shifts which
 are reflected by the genuine part of the loop contributions
 ${\rm Re}\Sigma_{gen}(s)$. These can be obtained by  a subtraction procedure
at a suitable mass scale $s_0$~\cite{Pennington:2007xr}:
\begin{eqnarray}
 {\rm Re}\Sigma_{gen}(s) ={\rm Re}\Sigma(s)-{\rm Re}\Sigma(s_0).\label{eq:gensigma}
\end{eqnarray}
Setting the scale $s_0$  needs a prescription. 
 It is reasonable to choose $s_0$ as the
 mass squared of the lowest lying bound state for a given quark flavor, which we take as
the $J^{PC}=0^{++}$ scalar tetraquark state. A different choice, 
namely $s_0=4m_{[bq]}^2$, will slightly modify our results and the effects caused by
the ambiguity in $s_0$ will be incorporated in estimating the systematic uncertainties
in our approach.  

Including the loop corrections,
 we now have  the following structure for the $2\times 2$ mass matrix
\begin{eqnarray}
M&=& \hat M+\sum_{i}c_{i}\bigg(\begin{array}{cc}\Gamma^{\tilde Z_b}_{i}&
  -\sqrt{\Gamma^{\tilde Z_b}_{i}\Gamma^{\tilde Z^\prime_b}_{i}}\\
  -\sqrt{\Gamma^{\tilde Z_b}_{i}\Gamma^{\tilde Z^\prime_b}_{i}}&
  \Gamma^{\tilde Z_b^\prime}_{i}\end{array}\bigg),
  \label{eq:MassMatrix}
\end{eqnarray}
where $i$ runs over the two-body channels shown in Fig.~\ref{fig:mesonloop}; the coefficients
 $c_i(s,s_0)$ are defined as 
 \begin{equation}
 	 c_i(s,s_0)\equiv -\frac{{\rm Re}\Sigma_{gen}(s)}{2\; {\rm Im}\Sigma(s) }\; , \label{eq:cidefinition}
 \end{equation} in which $s$ is taken as the
 physical mass squared from the data and ${\rm Re}\Sigma_{gen}(s)$ is determined as in Eq.~\eqref{eq:gensigma}.  The sign in the $\Upsilon(nS)\pi$ contributions to the off-diagonal
 terms is reversed due to the spin symmetry as shown in Eq.~\eqref{eq:Fierz}.  In the case of open bottom 
mesons,  the $B\bar B^*$ loop impacts on $M_{22}$ while $B^*\bar B^*$ 
modifies $M_{11}$.  Note, that
via the optical theorem the imaginary parts are directly related to the decay widths,
 and our parametrization
 in Eq.~\eqref{eq:MassMatrix} makes this manifest.

\begin{figure}
\begin{center}
\includegraphics[scale=0.65]{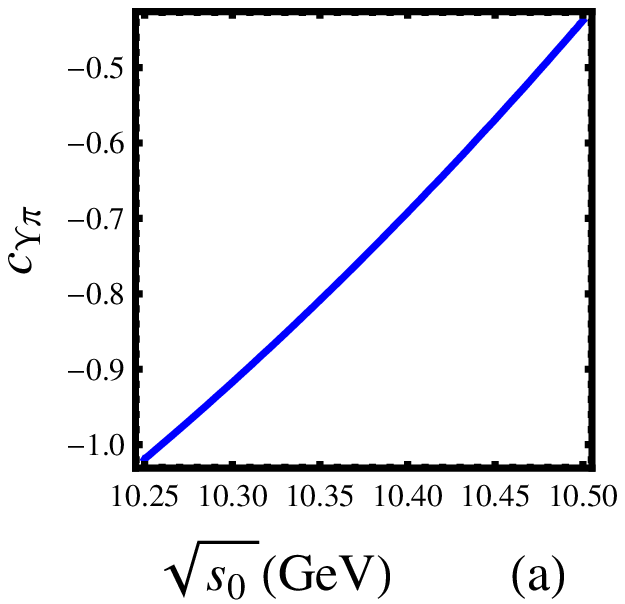}
\includegraphics[scale=0.65]{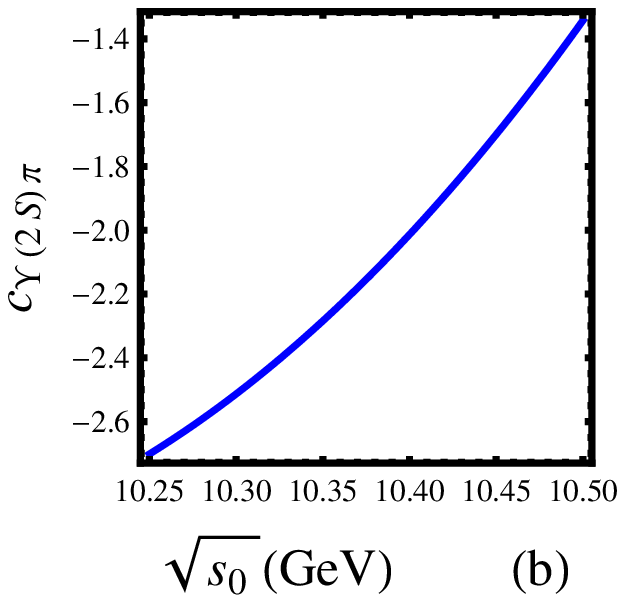}
\includegraphics[scale=0.65]{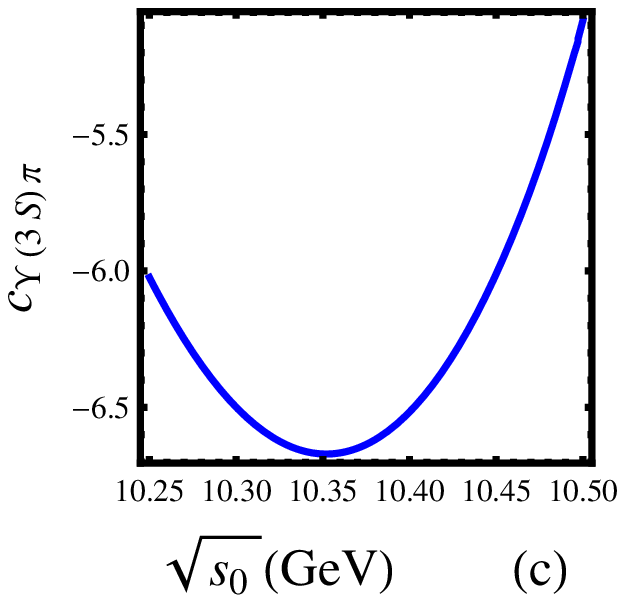}
\includegraphics[scale=0.65]{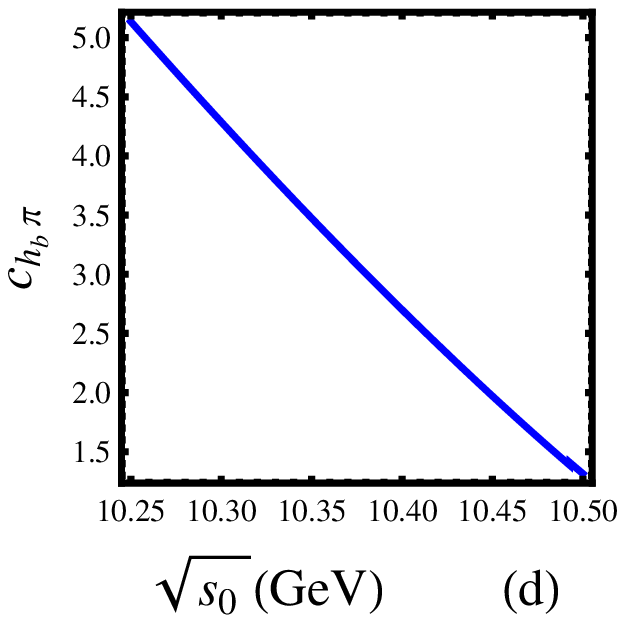}
\includegraphics[scale=0.65]{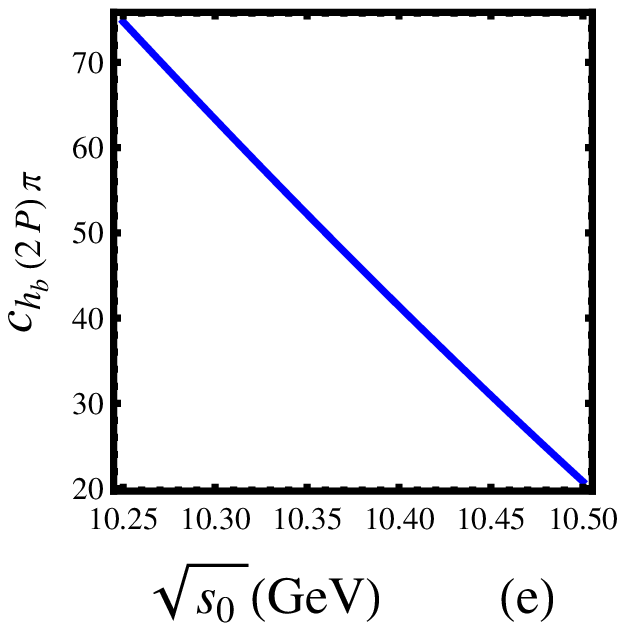}
\includegraphics[scale=0.65]{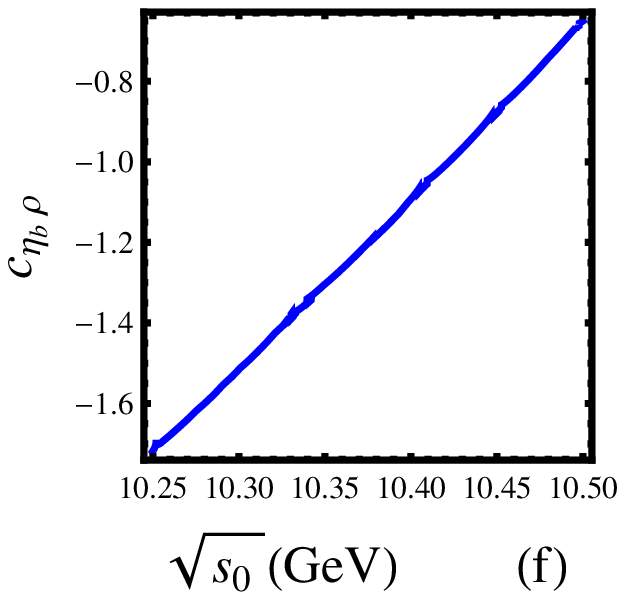}
\includegraphics[scale=0.65]{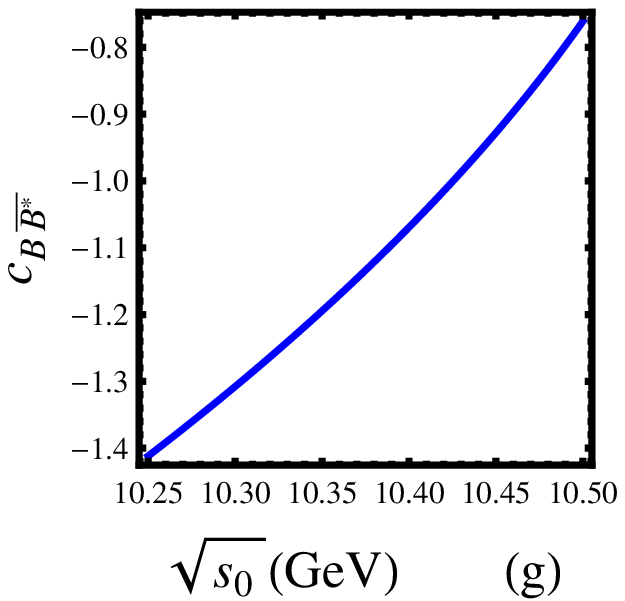}
\caption{Dependence of the coefficients $c_i$ on the subtraction scale $s_0$
calculated  with the Lagrangian in Eq.~\eqref{eq:VLagrangian}. The frames (a) to (g)
correspond to the intermediate states
 $i=\Upsilon(1S)\pi,
\Upsilon(2S)\pi,\Upsilon(3S)\pi, h_b(1P)\pi, h_b(2P)\pi, \eta_b\rho, B\overline{B^*}$,
 respectively.}
\label{fig:dependencescale}
\end{center}
\end{figure}

Choosing the subtraction point as $s_0 = [(10.385\pm 0.05) {\rm GeV}]^2$,
which corresponds to the mass of the lowest ($0^{++}$) tetraquark state
with a hidden $b\bar{b}$ quark content,  we 
estimate the following values for the coefficients
 $c_i$ (ignoring errors on the smaller $c_i$s):
 \begin{equation}
{\small
\begin{array}{|c|c|c|c|c|c|c}
c_{h_b(2P)\pi}
&
c_{\eta_b\rho}
&
c_{h_b(1P)\pi}
&
c_{B\bar B^*}
\\
\hline
45^{+11}_{-10}
& -1.1
& 3\pm 1
&-1.1
\end{array}.
}\nonumber
\end{equation}
For the analysis of $\Upsilon(nS)\pi$ contribution,  the Lagrangian
$
 {\cal L}= g_{V Z_b^{(\prime)}\pi} V^\mu Z_{b\mu}^{(\prime)} \pi 
$
with $V=Y_b,\Upsilon(nS)$  gives 
\begin{eqnarray}
 \Sigma(s)  &=&   -
\frac{g_{\tilde Z_b^{(\prime)} \Upsilon(nS)\pi}g_{\tilde Z_b^{(\prime)} \Upsilon(nS)\pi}^*}{(4\pi)^2}
\int_0^1 dx \Big[-  \log\big(\frac{\Lambda}{\mu^2}\big) \nonumber\\
&& + \frac{\Lambda}{2m_{\Upsilon(nS)}^2}\log\big(\frac{\Lambda}{\mu^2}\big)-\frac{\Lambda}{2m_{\Upsilon(nS)}^2}\Big],
\label{eq:sigmasUpi-1}
\end{eqnarray}
with $\Lambda= x^2 s-xs +xm_\pi^2 +(1-x) m_{\Upsilon(nS)}^2-i\epsilon$ and the relevant coefficients are predicted as
\begin{eqnarray}
c_{\Upsilon(1S)\pi}=-0.01,\;\; c_{\Upsilon(2S)\pi}=-0.1,\nonumber\\
 c_{\Upsilon(3S)\pi}=-1.3.\label{eq:Upi-Lagrangian1}
\end{eqnarray}  The expression for contributions from  the $Z_b^{(')}\to \eta_b\rho$ channel is similarly obtained
 by replacing the vector $\Upsilon(nS)$ by $\rho$ and the pseudoscalar $\pi$ by $\eta_b$.

 Instead of the Lagrangian specified above,  using the Lagrangian with the derivative of the pion field
  which is inspired by the chiral symmetry
\begin{eqnarray} 
 {\cal L}= g_{V Z_b^{(\prime)}\pi} V^\mu  i\buildrel\leftrightarrow\over\partial_\nu Z_{b\mu}^{(\prime)}  i\partial^\nu\pi, \label{eq:VLagrangian}
\end{eqnarray}
we have 
\begin{eqnarray}
 \Sigma(s)  &=&   -
\frac{g_{\tilde Z_b^{(\prime)} \Upsilon(nS)\pi}g_{\tilde Z_b^{(\prime)} \Upsilon(nS)\pi}^*}{(4\pi)^2}
\int_0^1 dx  \nonumber\\
&& \times \Bigg\{\frac{\Lambda^3}{4m_{\Upsilon(nS)}^2}\Big[4  \log\big(\frac{\Lambda}{\mu^2}\big)-5\Big] \nonumber\\
&& -\frac{\Lambda^2}{4m_{\Upsilon(nS)}^2}\Big[3  (4m_{\Upsilon(nS)}^2 +s(2-3x^2))\log\big(\frac{\Lambda}{\mu^2}\big)- \nonumber\\
&& - 8m_{\Upsilon(nS)}^2+ 12 s x^2 -7 s\Big] \nonumber\\
&& +\frac{\Lambda s}{2m_{\Upsilon(nS)}^2}\Big[[(8x^2-4)m_{\Upsilon(nS)}^2-  (x^2-1)^2s ]  \nonumber\\
&& +  [ (8-12x^2) m_{\Upsilon(nS)}^2 +s(1-x^2)^2]\log\big(\frac{\Lambda}{\mu^2}\big)\Big] \nonumber\\
&& - s^2(x^2-1)^2 \log\big(\frac{\Lambda}{\mu^2}\big)\Bigg\}~.
\label{eq:sigmasUpi-2}
\end{eqnarray}
This expression yields larger values for the coefficients $c_{\Upsilon(nS)\pi}$:
\begin{eqnarray}
c_{\Upsilon(1S)\pi}=-0.7,\;\; c_{\Upsilon(2S)\pi}=-2.1,\nonumber\\
 c_{\Upsilon(3S)\pi}=-6.5.
\end{eqnarray} 
It should be noted that  these numbers are much larger than the ones in Eq.~\eqref{eq:Upi-Lagrangian1}, due to the fact that the pion momentum  coming from the derivative in Eq.~\eqref{eq:VLagrangian} is small in the $Z_b^{(')}$ rest frame and thus suppresses the partial decay width and hence the denominator in the definition
 of $c_i$ as in Eq.~\eqref{eq:cidefinition}. 

The dependence of these coefficients on the subtraction scale  is shown in Fig.~\ref{fig:dependencescale},
 where the Lagrangian in Eq.~\eqref{eq:VLagrangian} has been used. The striking result is that the
coefficient $c_{h_b(2P)\pi}$ dominates by far all the others. 
The main reason for the dominance  of the coefficient $c_{h_b(2P)\pi}$ is that the
 limited  phase space and the p-wave decay character of $Z_b^{(\prime)}\to h_b(2P)\pi$ result
 in a comparably small value for the imaginary part of $\Sigma(s)$ compared to its real part. 
 In Ref.~\cite{Adachi:2011ji},  Belle collaboration has
reported the  measurements of the cross sections for $e^+e^-\to
\Upsilon(nS)\pi^+\pi^-$ and $e^+e^-\to h_b(mP)\pi^+\pi^-$ near the peak of the $\Upsilon(5S)$ resonance. Using the final state $\Upsilon(2S) \pi^+\pi^-$ as  normalization,
  they found that the ratios of the various cross sections are typically all of order 1. Thus,
 the partial widths for the different final states listed 
are comparable, which suggests a value of $O( 1 )$ MeV for each of them~\cite{Collaboration:2011pd}.
Thus, the  domination of the $h_b(2P)\pi$ channel in the meson-loop corrections
 to the $1^{+-}$ mass matrix is a consequence of this channel having the largest coefficient 
and the anticipated sizable partial decay width of $Z_b^{(\prime)}\to h_b(2P)\pi$. This is worked out quantitatively
later.

\begin{figure}
\begin{center}
\includegraphics[scale=0.6]{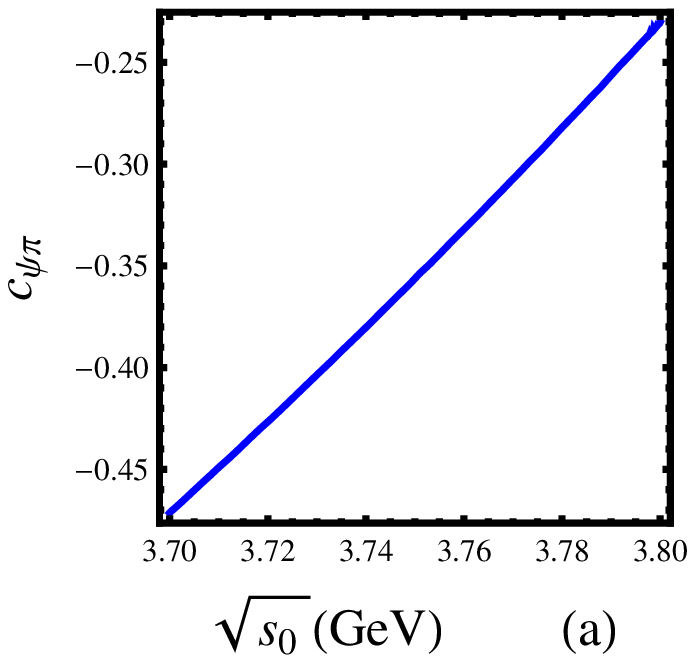}
\includegraphics[scale=0.6]{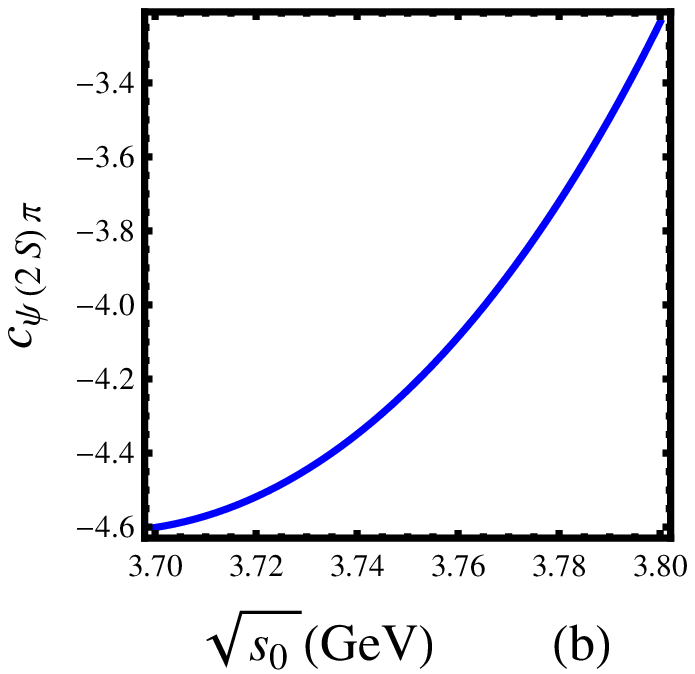} 
\includegraphics[scale=0.6]{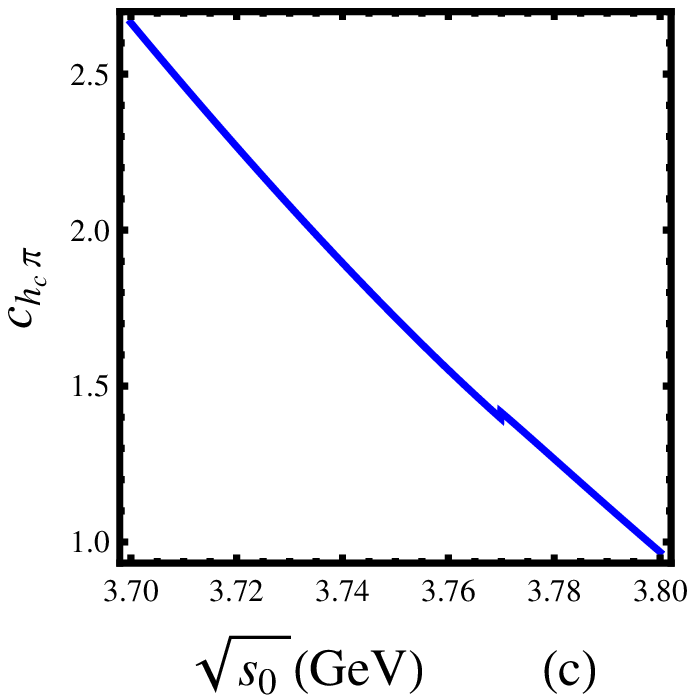} 
\includegraphics[scale=0.6]{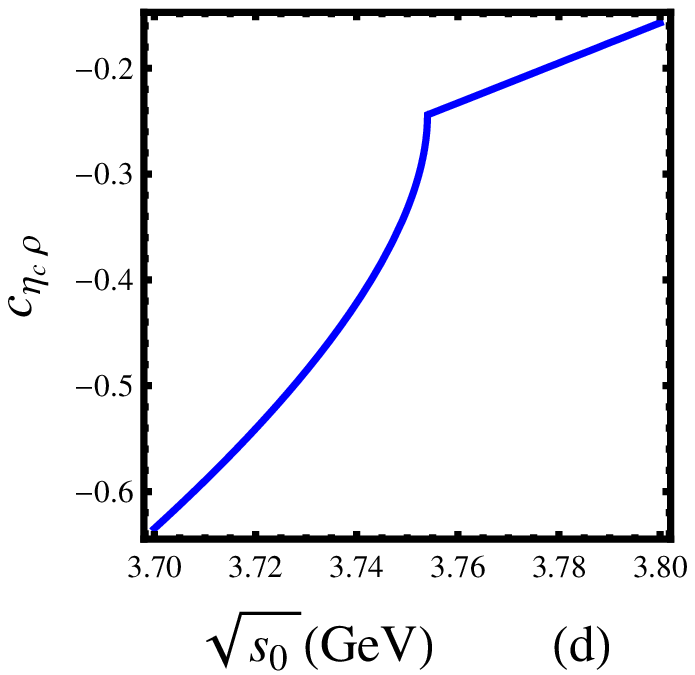}
\includegraphics[scale=0.6]{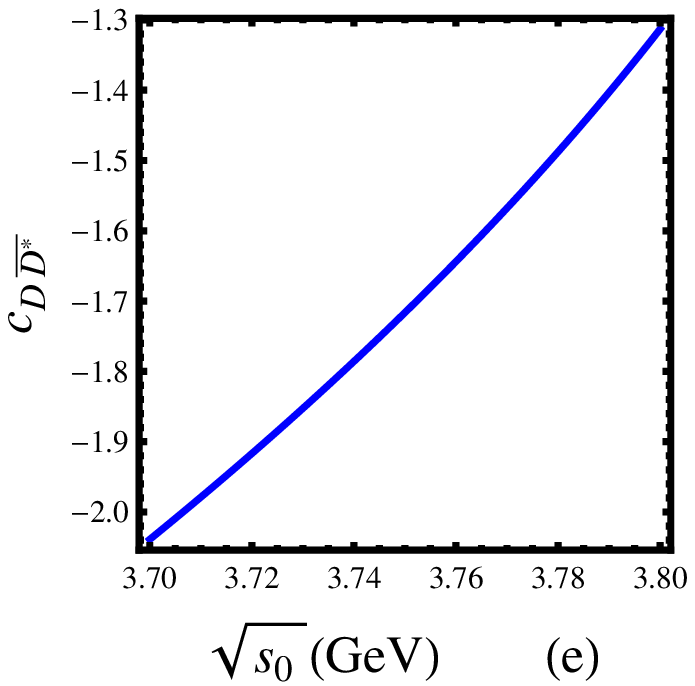}
\caption{Dependence of the coefficients $c_i$ for hadronic decay channels of the charmonium-like $Z_c^{(')}$ states on the subtraction scale $s_0$
calculated  with the Lagrangian in Eq.~\eqref{eq:VLagrangian}. The panels (a) to (e)
correspond to the intermediate states
 $i=J/\psi\pi,
\psi(2S)\pi,h_c\pi, \eta_c\rho, D\overline{D^*}$,
 respectively.}
\label{fig:dependencescale-charm}
\end{center}
\end{figure}

For comparison, we have  performed the same calculation for the hidden-charm tetraquark
 states whose masses are calculated in  the constituent diquark model by
Maiani et al.~\cite{Maiani:2004vq}, predicting  
  the  masses of the two $1^{+-}$ $c\bar{c}$ hidden tetraquark states as 
\begin{eqnarray}
 m_{Z_c}= 3.752 \;\; {\rm GeV},\;\; m_{Z_c'} = 3.882  \;\; {\rm GeV}.
\end{eqnarray} 
This yields a mass difference $\Delta m_{Z_c}=130$ MeV. Ignoring the isospin symmetry breaking
 effects, typically  a few MeV, the above estimates hold also for the 
 charged counterparts. 
Since the above masses  are very close to the estimate of the mass  of the lightest 
scalar $J^{PC}=0^{++}$ tetraquark state, $m_{S_c}= 3.723$ GeV~\cite{Maiani:2004vq},
the genuine meson-loop contributions, after subtraction, are expected to be small. 
We show the corresponding coefficients for various hadronic channels in
 Fig.~\ref{fig:dependencescale-charm}, where in order to
 determine the imaginary part the mass for the higher $1^{+-}$ state has been used as the
 physical mass. A striking difference between the coefficients shown in Figs. 2 and 3 is the
absence of the coefficient $c_{h_c(2P)\pi}$ in Fig.~3, as the decays $Z_c^{(\prime)} \to h_c(2P)\pi$
are not allowed kinematically.   Assuming that the partial decay widths in the
various channels $J/\psi \pi, \psi(2S)\pi, h_c(1P)\pi, \eta_c\rho$ etc. are of order $1$ MeV,
as in the decays of the $Z_b^{(\prime)}$, 
  we anticipate that the corrections due to the meson loops in the $Z_c$ - $Z_c^\prime$ mixing
are also typically  of the same order,
namely order 1  MeV, hence not significant. Thus, unlike the masses of the $Z_b$ -$Z_b^\prime$ complex,
 the masses for the hidden-charm tetraquarks calculated 
 in Ref.~\cite{Maiani:2004vq}  are not expected to be significantly modified by meson-loop
 effects. 

As already remarked, the $1^{+-}$ relatives of the $Z_b$ and $Z_b^\prime$ states in the charm sector,
$Z_c$ and $Z_c^\prime$, in our model 
 are expected to be produced in the decays of a $1^{--}$ hidden-charm tetraquark. The state $Y(4260)$ fits
 the bill. 
 The enhancement of the cross sections for $e^+e^-\to J/\psi\pi^+\pi^-$ and $e^+e^-\to h_c \pi^+\pi^-$
 seen by the CLEO collaboration at the center-of-mass energy around 4.26 GeV~\cite{CLEO:2011aa}
 is very likely a signature of their existence. In order to confirm or negate this scanrio,
 we suggest our experimental colleagues to scan over this mass region more precisely.

Returning to the discussion of the mass difference of the $1^{+}$ tetraquarks in the hidden $b\bar{b}$
sector, we note that it 
 is approximately given as 
$\Delta m_{Z_b}=  2\sqrt{ a^{\prime 2}+b^{\prime2}}$,
where $a^\prime= a- c_{i} (\Gamma^{\tilde Z_{b}}_{i}-\Gamma^{\tilde Z_{b}^\prime}_{i}) /2$,
 $b^\prime= b-c_{i} \sqrt{\Gamma^{\tilde Z_{b}}_{i} \Gamma^{\tilde Z_{b}^\prime}_{i} } $
and  $i$ denotes $h_b(2P)\pi$, as we keep only the dominant contribution.
The corresponding mass eigenstates are 
\begin{eqnarray}
 |Z_b\rangle =\cos\theta_{Z_b}|\tilde Z_b\rangle - \sin\theta_{Z_b} |\tilde Z_b^\prime\rangle,\nonumber\\
 |Z_b^\prime\rangle = \sin\theta_{Z_b}|\tilde Z_b\rangle +\cos\theta_{Z_b}|\tilde Z_b^\prime\rangle,
\end{eqnarray}
with  $\theta_{Z_b} = {\rm tan}^{-1} [ {b^\prime/ (a^\prime+ \Delta m_{Z_b} /{2})} ]$.

\begin{figure}
\begin{center}
\includegraphics[scale=0.55]{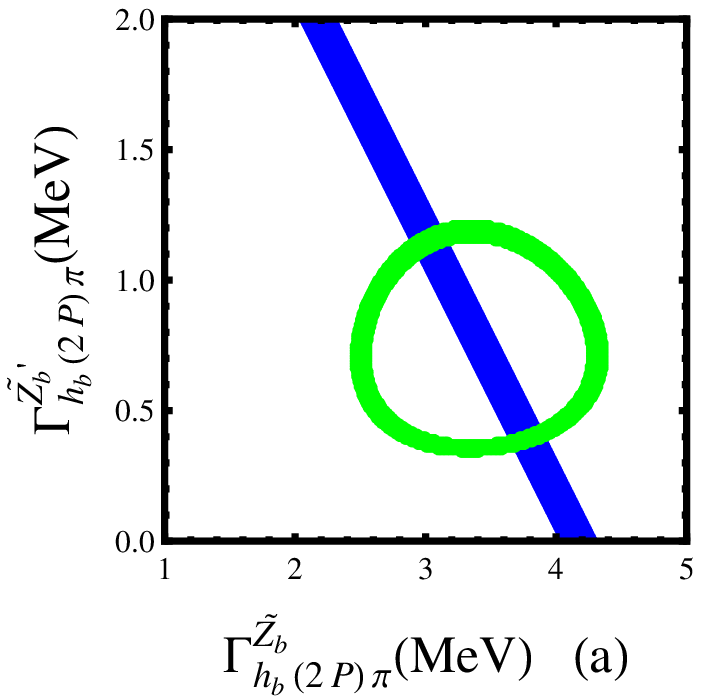}\hspace{0.5cm}
\raisebox{0.25cm}{\includegraphics[scale=0.5]{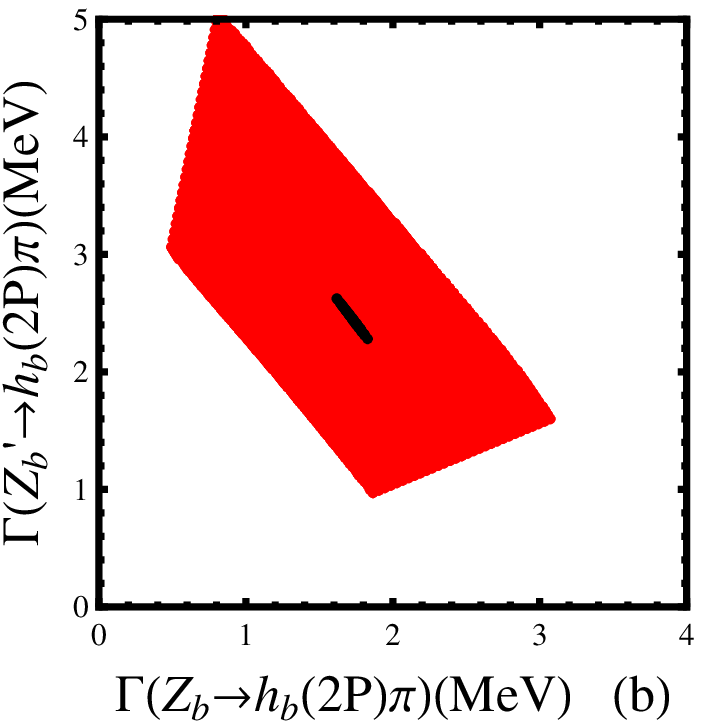}}
\caption{Constrained partial decay widths from the $Z_b$ and $Z_b^\prime$ masses measured by Belle.
 The left-hand panel shows the 
constraint on the partial decay widths of the tetraquark eigenstates $\tilde Z_b$ and $\tilde Z_b^\prime$.
 The circular (green) contour  is obtained by the mass difference $\Delta m_{Z_b}=45\pm 2.5$ MeV,
 while the slanted vertical (blue) band results  from the averaged mass
 $(m_{Z_b} + m_{Z_b^\prime})/2=10629.7\pm 2.5$ MeV for the default values
$\Delta=120$ MeV and $c_{h_b(2P)\pi}=45$.
 In the right-hand panel, the corresponding constraints on $Z_b$ and $Z_b^\prime$ partial decay
widths are depicted. The solid (black) region results from default values,
 while the  extended (red) region is obtained by varying these two parameters, 
as stated in the text. 
} \label{fig:contour}
\end{center}
\end{figure}

In Fig.~\ref{fig:contour},  we show the constrained partial decay widths from the masses 
observed by Belle.
The left panel shows the constraints on the widths of the tetraquark mass eigenstates
 $\tilde Z_b^{(\prime)}$  for the default values of $\Delta$ and
 $c_{h_b(2P)\pi}$.
In the spin-symmetry limit,  $\Gamma^{\tilde Z_{b}}_{i} $ and
$\Gamma^{\tilde Z_{b}'}_{i}$ are equal.
 As seen in this panel, the resulting contours
intersect at two regions,  the lower one of which implies large symmetry breaking effects
and hence is not entertained any further.  The upper region in which  the two couplings differ by
approximately $40\%$ is further analyzed. In the right-hand panel, the corresponding
 constraints on the $Z_b$ and $Z_b^\prime$ partial decay widths are depicted. The black region denotes
the default values of $\Delta$ and $c_i$ given above, while the
extended (red) region is obtained by 
the variations of these two parameters in the  ranges $\Delta=(120 \pm 30)$MeV and
$c_{h_b(2P)\pi}=45^{+11}_{-10}$. Based on this, we estimate
\begin{eqnarray}
&&\theta_{Z_b} =(-19^{+13}_{-17})^\circ, \nonumber\\
&&\Gamma(Z_b\to h_b(2P)\pi) =(1.7^{+1.3}_{-1.2}) {\rm MeV}, \nonumber\\
&&\Gamma(Z^\prime_b\to h_b(2P)\pi) =(2.5^{+2.5}_{-1.5}) {\rm MeV}.
\end{eqnarray}
We note that the mixing angle is small, implying that the mass eigenstates are close
to their respective tetraquark spin states. From the partial widths given above,
we extract the relative strength  of the coupling constants
\begin{eqnarray}
	r_{h_b(2P)\pi}\equiv|{g_{Z_b^\prime h_b(2P)\pi}}/{g_{Z_b h_b(2P)\pi}}|= 1.2^{+1.1}_{-0.5}.\label{eq:ratio}
\end{eqnarray}
In the Belle data~\cite{Collaboration:2011pd}, the ratio $r_{h_b(2P)\pi}$ is not measured directly; what is reported is the ratio $
a_i e^{i\phi_i} \equiv {g_{Y_b Z_b^\prime \pi} \times g_{Z_b^\prime i}}/({g_{Y_b Z_b \pi} \times g_{Z_bi}})$, which are
products of the production and the corresponding decay
amplitudes of the $Z_b$ and $Z_b'$ in the given final states. The updated value in~\cite{Collaboration:2011pd} is $a_{h_{b}(2P)\pi}=1.6^{+0.6+0.4}_{-0.4-0.6}$. 
An
analysis to estimate the relative amplitudes in all five final states reported in
Table I in~\cite{Collaboration:2011pd} is in progress  in the tetraquark context.
We anticipate that the couplings in the production
 amplitudes involving $Z_b$ and $Z_b^\prime$ are similar, i.e., $|g_{Y_b Z_b^\prime \pi}|\simeq
 |g_{Y_b Z_b\pi}|$ and hence $r_{h_b(2P)\pi}=a_{h_b(2P)\pi}$, in agreement with the Belle data.

Using the Lagrangian given in Eqs.~(\ref{eq:hbLagrangian}) and (\ref{eq:VLagrangian}), we have 
the following amplitudes for the decays
$Y_b \to \Upsilon(nS)\pi^+\pi^-$ and $Y_b \to h_b(mP)\pi^+\pi^-$  
\begin{eqnarray}
&& i \mathcal M(Y_b\to \Upsilon(nS) \pi^+\pi^-)
=
A_{nZ_b}+ i g_{Y_b Z_b \pi}  g_{Z_b \Upsilon(nS) \pi}   \nonumber\\
&&
\epsilon_{Y_b}  \cdot \epsilon^{*}_{\Upsilon(nS)} \big\{ ({\rm BW}_{Z_b}^{\Upsilon(nS)\pi^+} + a_{\Upsilon(nS)\pi} e^{i\phi_{\Upsilon(nS)\pi}}  {\rm BW}_{Z_b'}^{\Upsilon(nS)\pi^+}) \nonumber\\
&&   (m_{Y_b}^2- p_{\Upsilon(nS)\pi^+}^2) ( p_{\Upsilon(nS)\pi^+}^2-m_{\Upsilon(nS)}^2)  + (\pi^+\to \pi^-) \big\} ,\nonumber\\
&&i\mathcal M(Y_b\to h_b(mP) \pi^+\pi^-)
=
A_{nZ_b}^\prime-i g_{Y_b Z_b \pi}  g_{Z_b h_b \pi }\epsilon_{\mu\nu\alpha\beta}  \nonumber\\
&&
\epsilon_{Y_b}^\nu  p^\alpha_{h_b(mP)}  \epsilon^{*\beta}_{h_b(mP)}  \big\{
  p_{\pi^+}^\mu  (m_{Y_b}^2- p_{h_b(mP)\pi^+}^2) [{\rm BW}^{h_b(mP)\pi^+}_{Z_b}\nonumber\\
&&   + a_{h_b(mP)\pi} e^{i\phi_{h_b(mP)\pi}}  {\rm BW}^{h_b(mP)\pi^+}_{Z_b'}]+ (\pi^+\to \pi^-) \big\} ,
\end{eqnarray} 
with  
$
 {\rm BW}_{Z_b^{(\prime)}}^i =  
 [{p_{i} ^2 - m_{Z_b^{(\prime)}}^2 +i m_{Z_b^{(\prime)}} \Gamma_{Z_b^{(\prime)}} }]^{-1}.
$ 
Belle measurements show that the $a_i$s are roughly 1 within errors, while the phases
 $\phi_i$ are close to either $0$ or $180^\circ$, though the errors are rather large.  It is worth pointing out that the momentum dependence arising from the interaction Lagrangian given in
Eq.~(\ref{eq:VLagrangian}) 
are not  taken into account in the parametrization adopted by Belle.  Although the relative strength of the amplitudes, namely $a_ie^{i\phi_i}$,  is not affected,  the $\pi^\pm\Upsilon(nS)$ and $\pi^\pm h_b(mP)$ spectrum distributions  will be modified. 
$A_{nZ_b}$ and $A_{nZ_b}^\prime$ refer to the
non-$Z_b^{(\prime)}$ amplitudes in the indicated final states.  

The structure of 
$A_{nZ_b}$ was worked out in the tetraquark picture in great detail in
Refs.~\cite{Ali:2009es,Ali:2010pq}. As opposed to the amplitudes involved in typical dipionic 
heavy Quarkonia transitions, such as $\Upsilon(4S) \to \Upsilon(1S) \pi^+\pi^-$, which are modeled after
the Zweig-suppressed QCD multipole expansion~\cite{Brambilla:2010cs}, the amplitudes for the decays
$Y_b(10890) \to \Upsilon(nS) \pi^+\pi^-$ are not Zweig-forbidden, and hence they are
significant. In addition, they lead to a resonant structure in the $\pi \pi$
invariant mass spectrum. This is most marked in
the $\Upsilon(1S) \pi^+\pi^-$ mode in the form of the resonances $f_0(980)$ and
$f_2(1270)$. The measured dipionic invariant mass spectra by Belle in these final states is
in conformity with the predictions~\cite{Ali:2009es,Ali:2010pq}.  On the other hand,
the  amplitudes $A_{nZ_b}^\prime$ in the decays $Y_b(10890) \to h_b(mP)\pi^+\pi^-$ are
expected to be neither resonant nor numerically significant. Only the transition
  $Y_b(10890) \to h_b(1P) f_0(980)$ is marginally allowed, heavily suppressed by the
phase space and the $P$-wave decay character. The state $f_0(600)$ (or $\sigma(600)$)
contributes, in principle. However, as this is a very broad resonance, the higher mass part is
again suppressed by the phase space and hence the contribution of the $f_0(600)$
in the decay  $Y_b(10890) \to h_b(1P) f_0(600)$ is both small
and difficult to discern. This feature is also in accord with the Belle
data~\cite{Collaboration:2011pd}. Finally, we note that the absence of any anomalous
production of the states ($\Upsilon(nS) \pi^+\pi^-, h_b(mP)\pi^+\pi^-)$ in the decays of
 the bottomonium state
$\Upsilon(11020)$~\cite{Chen:2008xia} is anticipated in the tetraquark picture,
as opposed to the hadronic molecular interpretation of the $Z_b$ and $Z_b^\prime$ for which
the decays
 $\Upsilon(11020) \to Z_b^{(\prime)\pm}\pi^\mp \to \Upsilon(nS) \pi^+\pi^-, h_b(mP)\pi^+\pi^-$
 are expected to be enhanced by the
larger phase space compared to the corresponding decays from the $\Upsilon(5S)$.

In summary, we have presented a tetraquark interpretation of the two observed states
 $Z_b^\pm(10610)$ and $Z_b^\pm(10650)$. Combining the effective diquark-antidiquark Hamiltonian
with the meson-loop induced effects, we are able to account for the observed masses in terms of
the decay widths for the dominant channel $Z_b^{(\prime)\pm} \to h_b(2P) \pi^\pm$, obtaining a
ratio for the relative decay amplitudes in the decays $Z_b^{(\prime)\pm} \to h_b(mP) \pi^\pm$ which
 agrees with the Belle data. Together with the
resonant $\pi \pi$ structure in the decay modes $Y_b(10890) \to \Upsilon(nS) \pi^+\pi^-$,
first worked out in~\cite{Ali:2009es,Ali:2010pq}, this Letter provides additional support to the
tetraquark hypothesis involving the states $Y_b(10890)$,  $Z_b^\pm(10610)$ and $Z_b^\pm(10650)$. 
Precise spectroscopic measurements foreseen at the Super-B factories and at the LHC will provide
definitive answers to several issues raised here and will help resolve the current and long-standing
 puzzles in the exotic bottomonium sector.  

We acknowledge helpful discussions with Feng-kun Guo and Satoshi Mishima. W. W. is supported by
 the Alexander-von-Humboldt Stiftung.


\end{document}